Article

# Diffusion of Water Molecules on the Surface of Silica Nanoparticles—Insights from Nuclear Magnetic Resonance Relaxometry

Aleksandra Stankiewicz, Adam Kasparek, Elzbieta Masiewicz, and Danuta Kruk*

Read Online

ACCESS | Metrics & More | Article Recommendations | Supporting Information

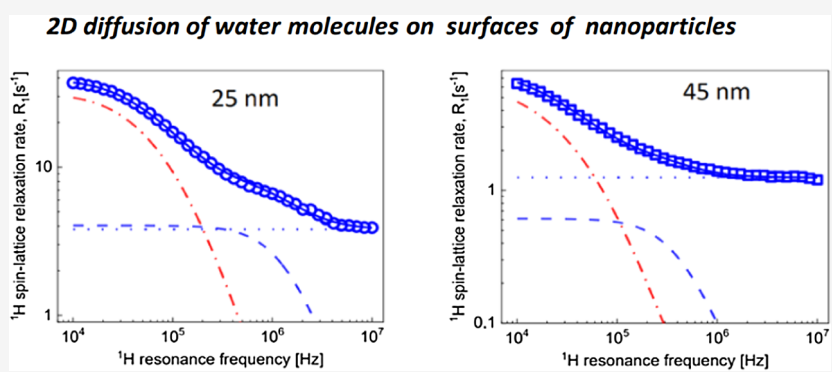

**ABSTRACT:** $^1$H spin-lattice nuclear magnetic resonance (NMR) relaxation experiments have been performed for water dispersions of functionalized silica nanoparticles of diameters of 25 and 45 nm. The experiments have been performed in a broad frequency range spanning 3 orders of magnitude, from 10 kHz to 10 MHz, versus temperature, from 313 to 263 K. On the basis of the data, two-dimensional translation diffusion (diffusion close to the nanoparticle surface within a layer of the order of a few diameters of water molecules) has been revealed. The translational correlation times as well as the residence life times on the nanoparticle surface have been determined. It has turned out that the residence lifetime is temperature-independent and is on the order of $5 \times 10^{-6}$ s for the smaller nanoparticles and by about a factor of 3 longer for the larger ones. The translational correlation time for the case of 25 nm nanoparticles is also temperature-independent and yields about $6 \times 10^{-7}$ s, while for the dispersion of the larger nanoparticles, the correlation times changed from about $8 \times 10^{-7}$ s at 313 K to about $1.2 \times 10^{-6}$ s at 263 K. In addition to the quantitative characterization of the two-dimensional translation diffusion, correlation times associated with bound water molecules have been determined. The studies have also given insights into the population of the bound and diffusing water on the surface water fractions.

## I. INTRODUCTION

The mechanism of molecular motion in the vicinity of surfaces is a subject of intensive investigations. Majority of the studies are theoretical, i.e., by means of molecular dynamics simulations.[1−4] Experimental means revealing the mechanism of molecular motion are very limited. With this respect, nuclear magnetic resonance (NMR) relaxometry is a unique method, revealing not only the time scale of dynamical processes but also their nature (mechanisms). For $^1$H nuclei, the dominating relaxation mechanism is provided by magnetic dipole−dipole interactions that fluctuate in time as a result of stochastic molecular motion (such as translation or rotation diffusion). "Classical" NMR experiments are performed at a single (high) magnetic field (resonance frequency). At a given resonance frequency, the most efficient relaxation pathway is associated with a dynamical process occurring on a time scale matching the reciprocal resonance frequency. Consequently, at high resonance frequencies, one probes only fast molecular motion.

Exploiting fast field cycling technology,[5,6] one can perform relaxation experiments in a broad range of resonance frequencies. Such studies are referred to as NMR relaxometry, and the spanned frequency range encompasses at least 3 orders of magnitude, from 10 kHz to 10 MHz (referring to the $^1$H resonance frequency). The broad frequency range enables probing dynamical processes on the time scale from milliseconds to nanoseconds in a single experiment: at low frequencies, the dominating relaxation contribution is associated with slow motion, with increasing frequency progressively faster dynamics comes into play. According to spin relaxation theory,[7−9] relaxation rates are given as a linear combination of spectral density functions that are Fourier transforms of the corresponding time correlation functions characterizing the motion causing stochastic fluctuations of the dipole−dipole interactions, the origin of the relaxation process. The mathematical form of the correlation function (hence the corresponding spectral density) depends on the mechanism of the motion.[9−16] In this way, NMR relaxometry possesses the unique potential to reveal the mechanism of the molecular motion (not only its time scale) from the shape of the frequency dependence of the spin−lattice relaxation rate (reciprocal relaxation time). The frequency dependence of the relaxation rate reflects the shape (the mathematical form) of the spectral density function—one can not only distinguish between rotational and translational dynamics[17,18] but also determine the dimensionality of the translation movement, i.e., distinguish between isotropic (three-dimensional), two-dimensional, and one-dimensional translation diffusion.[19−21] This advantage of NMR relaxometry has been exploited to reveal the mechanism of molecular and ionic motion in confinement in bulk and confinement.[19−30] In this context, one should point out two-dimensional (surface) diffusion of ions (ionic liquids) in the $SiO_2$ matrix[20] or two-dimensional diffusion of water molecules in systems including hyaluronic acid.[19] As far as water dynamics in systems including macromolecular fractions is concerned, recently, we have addressed the subject of water motion on the surface of proteins,[21] thoroughly testing different models of motion against $^1$H spin-lattice relaxation data for hydrated proteins. Spectral density functions characterizing translation diffusion of different dimensionalities have characteristic mathematical properties that enable unambiguous identification of the mechanism of motion. One of them is a linear dependence of the spin-lattice relaxation rates on squared root of the resonance frequency in the low-frequency range, when the condition $\omega\tau_{trans} < 1$ ($\omega$ denotes the resonance frequency in angular frequency units, while the quantity $\tau_{trans}$, referred to as a correlation time, describes the time scale of the translation diffusion) characteristic of free-dimensional (isotropic) translation diffusion;[10,14,31] this property stems from Taylor expansion of the spectral density function. Following this line, two-dimensional surface diffusion leads to a linear dependence of the spin-lattice relaxation rates on the logarithm of the resonance frequency, provided that conditions outlined in the next section are fulfilled.[15,16] Looking at the frequency dependencies of spin-lattice relaxation rates, one has to be aware that the overall relaxation rates result from contributions associated with several relaxation pathways (spin interactions) that are of intermolecular and intramolecular origin. The intermolecular interactions fluctuate in time as a result of mainly translation diffusion, while the intramolecular interactions are mediated by rotational and internal dynamics of the molecules. Consequently, the characteristic properties of the spectral density functions can be obscured by other relaxation contributions of a different shape—this subject has been addressed in our recent work devoted to water dynamics in the vicinity of the protein (bovine serum albumin) surface.[19]

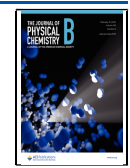

Despite the unique advantages of NMR relaxometry as a method providing insights into the mechanism of dynamical processes in molecular and ionic systems, such studies are rarely reported in the literature, mainly because of the limited availability of this kind of NMR equipment and the theoretical challenges of the data analysis. In this work, we address the subject of water dynamics in solutions of diamagnetic, silica (triethoxylpropylaminosilane-functionalized) nanoparticles. The objects are relatively large; we consider nanoparticles of diameters of 25 and 45 nm. The purpose of the work is twofold. The first goal is to reveal the mechanism of water motion in systems including relatively large nano-objects with focus on the influence of the nanoparticle size on the water dynamics, while the second goal is methodological: using this example, we shall present the theoretical approach enabling the identification of the mechanism of molecular motion in the vicinity of surfaces.

## II. THEORY

$^1$H spin-lattice relaxation processes are primarily caused by $^1$H−$^1$H magnetic dipole−dipole interactions. According to the spin relaxation theory,[7−9] the spin-lattice relaxation rate, $R_1(\omega)$ ($\omega$ denotes the resonance frequency in angular frequency units), is given as

$$R_1(\omega) = C_{DD}[J(\omega) + 4J(2\omega)] \quad (1)$$

where $J(\omega)$ denotes a spectral density function, Fourier transform of the corresponding time correlation function characterizing stochastic fluctuations of the spin interactions causing the relaxation process, and $C_{DD}$ denotes a dipolar relaxation constant determined by the amplitude of the dipole−dipole coupling. For exponential correlation functions, the spectral density takes the form of a Lorentzian function. Consequently, the relaxation rate is given as[7−9]

$$R_1(\omega) = C_{DD}\left[\frac{\tau_c}{1 + (\omega\tau_c)^2} + \frac{4\tau_c}{1 + (2\omega\tau_c)^2}\right] \quad (2)$$

where $\tau_c$ denotes a correlation time, a characteristic time constant describing the time scale of the molecular motion leading to the stochastic fluctuations of the dipole−dipole interactions causing the relaxation process. Dipole−dipole interactions can be of intramolecular or intermolecular origin. Consequently, there are several contributions to the overall relaxation process (several relaxation pathways) that are associated with different dynamical processes. The mathematical form of the correlation function (and hence the spectral density function) depends on the mechanism of motion (as already pointed out for the exponential correlation function and the corresponding Lorentzian spectral density). In the case of two-dimensional translation diffusion (diffusion close to the nanoparticle surface within a layer of the order of a few diameters of water molecules), the spectral density function, $J_{2D}(\omega)$, takes the form[13−20]

$$J_{2D}(\omega) = \tau_{trans} \ln\left[\frac{1 + (\omega\tau_{trans})^2}{\left(\frac{\tau_{trans}}{\tau_{res}}\right)^2 + (\omega\tau_{trans})^2}\right] \quad (3)$$

where $\tau_{trans}$ denotes the translational correlation time defined as $\tau_{trans} = \frac{d^2}{2D_{trans}}$,[10,11,31] $D_{trans}$ is the translation diffusion coefficient, and $d$ is referred to as a distance of the closest approach of the interacting nuclei. In the case of spherical molecules with nuclei (spins) placed in the molecular center, the distance of the closest approach is equal to a sum of the molecular radii (that gives the molecular diameter for identical

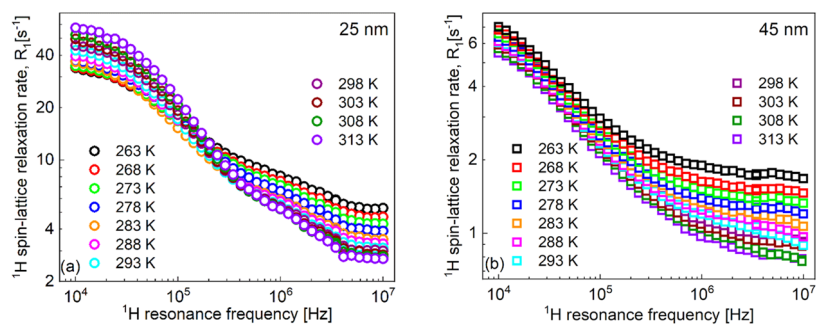

**Figure 1.** $^1$H spin-lattice relaxation rates for water dispersions of silica nanoparticles functionalized with triethoxylpropylaminosilane: (a) 25 nm diameter and $2.44 \times 10^{-2}$ mmol/dm$^3$ and (b) 45 nm diameter and $4.62 \times 10^{-3}$ mmol/dm$^3$.

molecules). The quantity $\tau_{res}$ is referred to as a residence lifetime on the surface.

Following eq 1, the relaxation rate associated with dipole–dipole interactions modulated by two-dimensional translation diffusion is given as

$$R_1(\omega) = C_{trans}\tau_{trans}\left[\ln\left[\frac{1+(\omega\tau_{trans})^2}{\left(\frac{\tau_{trans}}{\tau_{res}}\right)^2+(\omega\tau_{trans})^2}\right]\right.$$

$$\left.+ 4\ln\left[\frac{1+(2\omega\tau_{trans})^2}{\left(\frac{\tau_{trans}}{\tau_{res}}\right)^2+(2\omega\tau_{trans})^2}\right]\right] \quad (4)$$

where $C_{trans}$ denotes the corresponding dipolar relaxation constant. When $\frac{\tau_{trans}}{\tau_{res}} < \omega\tau_{trans}$, i.e., when $\omega\tau_{res} > 1$, eq 3 converges to the simplified form

$$J_{2D}(\omega) = \tau_{trans}\ln[1+(\omega\tau_{trans})^{-2}] \quad (5)$$

Following this line, when $\omega\tau_{res} > 1$ and $\omega\tau_{trans} \ll 1$, the spectral density (hence the corresponding relaxation rate) shows a linear dependence on $\ln\omega$.

It is expected that in the broad frequency range covered in NMR relaxometry experiments, the overall relaxation rate includes several relaxation contributions originating from intermolecular and intramolecular dipole–dipole interactions. It turned out that the results presented in this work can be reproduced in terms of three relaxation contributions

$$R_1(\omega) = C_{trans}\tau_{trans}\left[\ln\left[\frac{1+(\omega\tau_{trans})^2}{\left(\frac{\tau_{trans}}{\tau_{res}}\right)^2+(\omega\tau_{trans})^2}\right]\right.$$

$$\left.+ 4\ln\left[\frac{1+(2\omega\tau_{trans})^2}{\left(\frac{\tau_{trans}}{\tau_{res}}\right)^2+(2\omega\tau_{trans})^2}\right]\right]$$

$$+ C_{DD}\left[\frac{\tau_c}{1+(\omega\tau_c)^2}+\frac{4\tau_c}{1+(2\omega\tau_c)^2}\right] + A \quad (6)$$

The first term corresponds to a relaxation contribution associated with two-dimensional translation diffusion of the solvent (water) molecules on the nanoparticle surface, the second one describes a relaxation process expressed in terms of Lorentzian spectral densities (exponential correlation functions), it can be attributed to the dynamics of solvent molecules attached to its surface, and the third one, the frequency-independent term, $A$, corresponds to a relaxation process associated with a motion occurring on the time scale of tenths of nanoseconds or faster, so in the covered frequency range, the condition $\omega\tau < 1$ (where $\tau$ denotes the corresponding correlation time) is fulfilled.

### III. MATERIALS AND METHODS

$^1$H spin-lattice NMR relaxation experiments have been performed for water dispersions of two kinds of silica (triethoxylpropylaminosilane-functionalized) nanoparticles in the frequency range from 10 kHz to 10 MHz versus temperature from 263 to 313 K (below 263 K, the systems have frozen). The dispersions have been purchased from Sigma-Aldrich. The nanoparticles differ in the size: for the first kind, the diameter is 25 nm and for the second one, the diameter is 45 nm (obtained by dynamic light scattering according to Sigma-Aldrich). Taking into account the solid content of the dispersions that yields 27.5% for the 25 nm nanoparticles and 29.1% for the 45 nm nanoparticles and density of the dispersions provided by Sigma-Aldrich of 1.158 g/cm$^3$ for the case of 25 nm nanoparticles and 1.2 g/cm$^3$ for the case of 45 nm nanoparticles, the concentrations of the nanoparticles have been estimated as $2.44 \times 10^{-2}$ and $4.62 \times 10^{-3}$ mmol/dm$^3$ for the mixtures of 25 and 45 nm nanoparticles, respectively.

A NMR relaxometer Stelar s.r.l. (Mede (PV), Italy) with a temperature accuracy of 1 K was used for performing the $^1$H spin-lattice relaxation experiments. For resonance frequencies lower than 4 MHz, prepolarization at the magnetic field of 0.19 T (that corresponds to the resonance frequency of 8 MHz) was applied. The switching time of the magnet was set to 3 ms, with a recycle delay and polarization time being 5 times longer than the spin-lattice relaxation time. The $^1$H magnetization values were recorded versus time using a logarithmic time scale for 32 acquisitions. The relaxation processes turned out to be single-exponential in all cases. Examples of the $^1$H magnetization curves (magnetization versus time) are shown in the Supporting Information (Figures S1 and S2).

### IV. RESULTS AND ANALYSIS

Figure 1 shows $^1$H spin-lattice relaxation data for water dispersions of the two kinds of nanoparticles versus temperature.

The concentrations of the nanoparticles are different. We have rescaled the relaxation rates to 1 mmol/dm$^3$ concen-tration. Before doing that (dividing by $2.44 \times 10^{-2}$ and $4.62 \times 10^{-3}$ for the case of 25 and 45 nm nanoparticles, respectively), relaxation rates for bulk water have been subtracted. The rescaled data are shown in Figure 2.

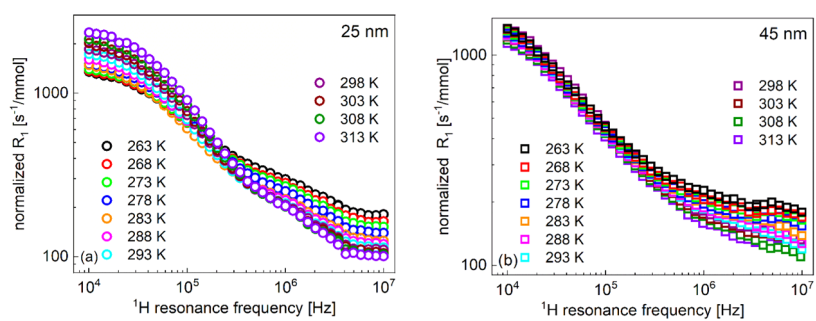

**Figure 2.** $^1$H spin-lattice relaxation rates for water dispersions of silica nanoparticles normalized to the concentration of 1 mmol/dm$^3$: (a) 25 nm diameter and (b) 45 nm diameter. The relaxation rates for bulk water, 0.24 s$^{-1}$ for 313 K, 0.26 s$^{-1}$ for 308 K, 0.28 s$^{-1}$ for 303 K, 0.31 s$^{-1}$ for 298 K, 0.34 s$^{-1}$ for 293 K, 0.38 s$^{-1}$ for 288 K, 0.43 s$^{-1}$ for 283 K, 0.49 s$^{-1}$ for 278 K, 0.57 s$^{-1}$ for 273 K, 0.68 s$^{-1}$ for 268 K, and 0.85 s$^{-1}$ for 263 K, have been subtracted before the normalization.

The normalization has been solely performed for a more straightforward comparison of the relaxation data, as there is no proof that the relaxation rates are indeed proportional to the concentration. Nevertheless, the comparison shows that while the relaxation rates are similar (under the assumption of proportionality to the concentration), the shapes of the frequency dependencies of the spin−lattice relaxation rates are considerably different. The shape of the relaxation data for the case of 25 nm nanoparticles clearly anticipates the presence of three relaxation contributions. The data have been analyzed in terms of eq 6 with six adjustable parameters: $C_{trans}$, $\tau_{trans}$, $\tau_{res}$, $C_{DD}$, $\tau_c$, and $A$. The intention of the analysis has been to keep the dipolar relaxation constants $C_{DD}$ and $C_{trans}$ temperature-independent. This attempt has failed; it has turned out that to reproduce the relaxation data, it is necessary to allow for changes in $C_{trans}$ with temperature. At the same time, the correlation times $\tau_{trans}$ and $\tau_{res}$ show a very weak temperature dependence; in practice, the quantities can be treated as temperature-independent. The parameters obtained for the dispersion, including 25 nm nanoparticles, are collected in Table 1.

The dipolar relaxation constant associated with the translation motion decreases with decreasing temperature by a factor of about 2 in the temperature range from 313 to 273 K, while the frequency-independent term, $A$, decreases with increasing temperature. The correlation time, $\tau_c$, ranges from $7.83 \times 10^{-8}$ s at 273 K (and below at 268 and 263 K) to $4.86 \times 10^{-8}$ s at 313 K. In Figure 3a−d, examples of the overall fits decomposed into the individual relaxation contributions are shown; the results for the remaining temperatures are included in the Supporting Information (Figure S3).

The same strategy was applied to the data obtained for the dispersion of the 45 nm nanoparticles. In this case, both dipolar relaxation constants, $C_{DD}$ and $C_{trans}$, vary with temperature. The only temperature-independent quantity is the residence lifetime, $\tau_{res}$. The parameters obtained for this dataset are listed in Table 2.

One should note that the relaxation constant $C_{DD}$ decreases from $7.23 \times 10^5$ to $5.56 \times 10^5$ Hz$^2$ in the temperature range from 288 to 278 K, while the relaxation constant $C_{trans}$ decreases from $2.41 \times 10^5$ to $2.11 \times 10^5$ Hz$^2$ in the range from 298 to 288 K. The temperature changes of the dipolar relaxation constants can reflect exchange effects being more significant at lower temperatures.[15]

Examples of relaxation data analysis are shown in Figure 4. The decomposition of the overall relaxation into the individual contributions for the remaining temperatures is shown in the Supporting Information (Figure S4).

Before discussing the obtained results, it is worthy to focus on the conditions required for observing a linear dependence of the relaxation rate, $R_1$, on $\sqrt{\omega}$: $\omega\tau_{trans} < 1$ and $\omega\tau_{res} > 1$. For the parameters obtained for the case of 25 nm nanoparticles, $\tau_{trans} = 5.84 \times 10^{-7}$ s and $\tau_{res} = 4.54 \times 10^{-6}$ s, one obtains $\omega\tau_{trans} < 1$ for the resonance frequency below approximately $2.7 \times 10^5$ Hz, while the condition $\omega\tau_{res} > 1$ is fulfilled for the resonance frequency above approximately $3.5 \times 10^4$ Hz. The vertical lines in Figure 5a indicate the range from $3.0 \times 10^4$ to $2.0 \times 10^5$ Hz; the linear dependence for the data at 313 K is shown. One can also see deviations from the linearity outside this range. Analogously, for the case of 45 nm nanoparticles, the expected linearity begins above frequencies about $1 \times 10^4$ Hz (this corresponds to the condition $\omega\tau_{res} > 1$ for $\tau_{res} = 1.43 \times 10^{-5}$ s) and ends below approximately $1.5 \times 10^5$ Hz (this corresponds to the condition $\omega\tau_{trans} < 1$ for $\tau_{trans} = 1 \times 10^{-6}$ s; $\tau_{trans}$ varies from $7.80 \times 10^{-7}$ to $1.18 \times 10^{-6}$ s and the value of $1 \times 10^{-6}$ s captures the average). This implies that the linear dependence is expected to begin already at the low frequency limit of the experiment (that is, $1 \times 10^4$ Hz). This is confirmed in Figure 5b, which shows the linearity for the case of 45 nm nanoparticles. The vertical line corresponds to the frequency of $1.5 \times 10^5$ Hz. The linear fit for the data at 263 K confirms the linear dependence in the range from $1 \times 10^4$ to $1 \times 10^5$ Hz and deviations above that frequency.

**Table 1. Parameters Obtained from the Analysis of $^1$H Spin-Lattice Relaxation Data for Water Dispersion of 25 nm Silica Nanoparticles**[a]

| temp. [K] | $\tau_c$ [$\times 10^{-8}$ s] | $C_{trans}$ [$\times 10^6$ Hz$^2$] | $A$ [s$^{-1}$] |
|---|---|---|---|
| 313 | 4.86 ± 0.46 | 4.63 ± 0.50 | 2.43 ± 0.14 |
| 308 | 4.86 ± 0.46 | 4.21 ± 0.42 | 2.50 ± 0.14 |
| 303 | 4.86 ± 0.47 | 3.85 ± 0.40 | 2.63 ± 0.16 |
| 298 | 4.90 ± 0.47 | 3.55 ± 0.37 | 2.75 ± 0.17 |
| 293 | 4.94 ± 0.48 | 3.27 ± 0.35 | 2.89 ± 0.18 |
| 288 | 5.17 ± 0.50 | 2.97 ± 0.29 | 3.11 ± 0.21 |
| 283 | 5.22 ± 0.49 | 2.66 ± 0.27 | 3.30 ± 0.25 |
| 278 | 6.56 ± 0.56 | 2.61 ± 0.27 | 3.80 ± 0.27 |
| 273 | 7.83 ± 0.65 | 2.25 ± 0.24 | 4.22 ± 0.31 |
| 268 | 7.83 ± 0.65 | 2.25 ± 0.24 | 4.27 ± 0.31 |
| 263 | 7.83 ± 0.67 | 2.25 ± 0.24 | 4.60 ± 0.34 |

[a]$C_{DD} = (1.23 \pm 0.16) \times 10^7$ Hz$^2$, $\tau_{trans} = (5.84 \pm 0.71) \times 10^{-7}$ s, and $\tau_{res} = (4.54 \pm 1.32) \times 10^{-6}$ s.

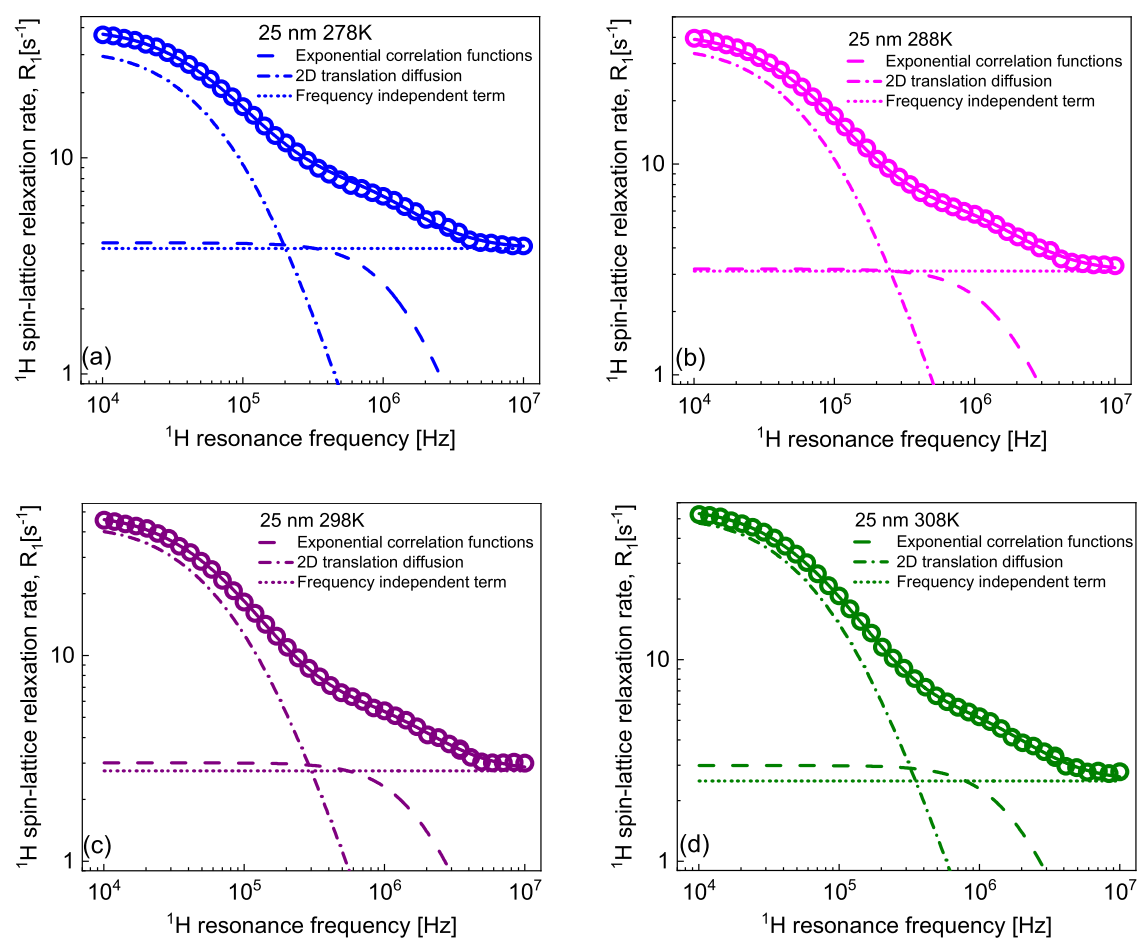

**Figure 3.** (a–d) $^1$H spin-lattice relaxation rates for water dispersion of silica nanoparticles of 25 nm diameter and $2.44 \times 10^{-2}$ mmol/dm$^3$ concentration. Solid lines, fits in terms of eq 6 decomposed into a relaxation contribution associated with two-dimensional translation diffusion (dashed-dotted lines), a relaxation contribution expressed in terms of Lorentzian spectral densities (dashed lines), and a frequency-independent term (dotted lines).

**Table 2. Parameters Obtained from the Analysis of $^1$H Spin-Lattice Relaxation Data for Water Dispersion of 45 nm Silica Nanoparticles**[a]

| temp. [K] | $C_{DD}$ [×10$^5$ Hz$^2$] | $\tau_c$ [×10$^{-7}$ s] | $C_{trans}$ [×10$^5$ Hz$^2$] | $\tau_{trans}$ [×10$^{-7}$ s] | $A$ [s$^{-1}$] |
|---|---|---|---|---|---|
| 313 | 7.23 ± 0.31 | 1.24 ± 0.15 | 2.41 ± 0.12 | 7.80 ± 0.64 | 0.81 ± 0.13 |
| 308 | 7.23 | 1.40 ± 0.17 | 2.41 | 8.02 ± 0.69 | 0.84 ± 0.11 |
| 303 | 7.23 | 1.45 ± 0.17 | 2.41 | 8.45 ± 0.70 | 0.91 ± 0.21 |
| 298 | 7.23 | 1.73 ± 0.22 | 2.41 ± 0.16 | 10.42 ± 0.84 | 0.99 ± 0.23 |
| 293 | 7.23 | 1.78 ± 0.22 | 2.14 ± 0.15 | 10.64 ± 0.84 | 1.01 ± 0.17 |
| 288 | 7.23 ± 0.45 | 1.86 ± 0.23 | 2.11 ± 0.11 | 11.80 ± 0.95 | 1.07 ± 0.11 |
| 283 | 6.71 ± 0.41 | 2.10 ± 0.25 | 2.11 ± 0.13 | 11.80 ± 0.91 | 1.17 ± 0.13 |
| 278 | 5.56 ± 0.33 | 2.19 ± 0.25 | 2.11 | 11.80 | 1.25 ± 0.21 |
| 273 | 5.56 ± 0.30 | 2.19 ± 0.21 | 2.11 | 11.80 | 1.38 ± 0.23 |
| 268 | 5.56 | 2.19 | 2.11 | 11.80 | 1.51 ± 0.22 |
| 263 | 5.56 | 2.19 | 2.11 | 11.80 | 1.74 ± 0.20 |

[a]$\tau_{res} = (1.43 \pm 0.23) \times 10^{-5}$ s. The quantities that do not include uncertainties were fixed in the fits.

The experimental verification of the linear dependence of the spin−lattice relaxation rates on the logarithm of the resonance frequency confirms the assumption of two-dimensional translation diffusion of water molecules on the nanoparticle surfaces.

## V. DISCUSSION

The analysis of the $^1$H spin−lattice relaxation data has revealed the two-dimensional mechanism of water diffusion on the surface of silica nanoparticles functionalized with triethoxylpropylaminosilane. It has turned out that the correlation times (hence the translation diffusion coefficient) are very weakly dependent on temperature in the covered temperature range from 263 to 313 K. For the case of the nanoparticle diameter of 25 nm, the correlation time $\tau_{trans}$ yields $5.84 \times 10^{-7}$ s and can be treated as temperature-independent. This corresponds to the translation diffusion coefficient $D_{\text{trans}} = 6.24 \times 10^{-14}$ m$^2$/s (calculated from the relationship $\tau_{\text{trans}} = \frac{d^2}{2D_{\text{trans}}}$,[10,11,31] where $d = 2.7$ Å denotes the diameter of the water molecule). For the larger nanoparticles (45 nm), the translational correlation time at 313 K is somewhat longer ($\tau_{\text{trans}} = 7.80 \times 10^{-7}$ s) than for the 25 nm nanoparticles; this corresponds to the translation diffusion coefficient $D_{\text{trans}} = 4.67 \times 10^{-14}$ m$^2$/s. The correlation time becomes progressively longer with temperature, reaching $1.18 \times 10^{-6}$ s ($D_{\text{trans}} = 3.09 \times 10^{-14}$ m$^2$/s) at 298 K and then remaining unchanged with decreasing temperature. The ratio between the diffusion coefficients for the case of 45 nm nanoparticles at 313 and 263 K is about 1.5 (so the temperature dependence is weak), while the ratio between the diffusion coefficients at 263 K for 45 and 25 nm nanoparticles yields 2.0. This comparison shows that the translation dynamics of water molecules is similar for both kinds of the nanoparticles. Slowing down of translation diffusion on surfaces in porous materials was reported.[15] Analogous effects were reported for protein surfaces.[24] In both cases, the residence lifetime (the time during which water molecules stay attached to the nanoparticle surface between the subsequent diffusion jumps) is temperature-independent. For the smaller nanoparticles, it yields $\tau_{res}$ = 4.54 × 10$^{-6}$ s, while for the larger ones, it is longer by a factor of about 3 (1.43 × 10$^{-5}$ s). One could expect that the values will be more close; however, the different shapes of the frequency dependencies of the relaxation rates for the dispersions of the 25 and 45 nm nanoparticles (Figures 1 and 5) clearly indicate differences in the residence lifetime. The relaxation constant $C_{trans}$ for the case of 25 nm nanoparticles varies from 4.63 × 10$^6$ Hz$^2$ at 313 K to 2.25 × 10$^6$ Hz$^2$ at 263 K. Although the difference is not large (factor of 2), one rather expects this parameter to be temperature-independent. The relaxation constant stems, however, from intermolecular dipole−dipole interactions (interactions between $^1$H nuclei of different water molecules) and it depends on the effective distance between the hydrogen atoms of neighboring molecules diffusing in the vicinity of the nanoparticle surface (like $r^{-3}$), where $r$ denotes the distance. Some changes in the arrangement of the molecules on the surface can lead to changes in the relaxation constant. For the larger nanoparticles, the relaxation constant is almost temperature-independent—it varies from 2.41 × 10$^5$ Hz$^2$ at 313 K to 2.11 × 10$^5$ Hz$^2$ at 263 K. We shall come back to the subject of the dipolar relaxation constants below. The correlation time $\tau_c$ can be attributed to water molecules attached to the surface. For the smaller nanoparticles, the values vary from 4.86 × 10$^{-8}$ s at 313 K to 7.83 × 10$^{-8}$ s at 268 K—the changes with temperature do not exceed a factor of 2. For the larger nanoparticles, the correlation time is longer (as expected), from 1.24 × 10$^{-7}$ s at 313 K to 2.19 × 10$^{-7}$ s at 268 K, with also a weak temperature dependence. Coming back to the values of the dipolar relaxation constants, the ratio between the concentrations of the 25 and 45 nm nanoparticles is about 5 (5.28). Rescaling the dipolar relaxation constants for the dispersion of the 45 nm nanoparticles, one obtains the following: rescaled $C_{trans}$ varies from 1.27 × 10$^6$ Hz$^2$ (313 K) to 1.11 × 10$^6$ Hz$^2$ (263 K), while the corresponding values for the dispersion of the 25 nm nanoparticles are 4.63 × 10$^6$ Hz$^2$ (313 K) and 2.25 × 10$^6$ Hz$^2$ (263 K), and the ratio ranges from about 4 (313 K) to about 2 (263 K). For the intramolecular dipolar relaxation constant, $C_{DD}$, one obtains after rescaling the values for the case of 45 nm nanoparticles of 3.82 × 10$^6$ Hz$^2$ (313 K) and 2.94 × 10$^6$ Hz$^2$ (263 K), compared to 1.23 × 10$^7$ Hz$^2$ for the 25 nm nanoparticles; this gives the ratio from about 3 (313 K) to about 4 (263 K). The dipolar relaxation constant, $C_{DD}$, is proportional to the mole fraction of the bound water molecules.[17] This might indicate that the fraction of bound water molecules is lower for the larger nanoparticles (referring to the same concentration of both kinds of the nanoparticles); however, one should be aware that exchange processes between water fractions might affect (to a certain extent) the effective value of $C_{DD}$.[15,17] This is also reflected by the relaxation constant $C_{trans}$ that depends on the ratio between the number of water molecules diffusing close to the nanoparticle surface within a layer on the order of a few diameters of water molecules and the bulk water population. At the higher temperatures, the ratio also reaches about 4; the decrease with temperature resulting from the decrease of $C_{trans}$ for the case of 25 nm nanoparticles can be caused (as already pointed out) by even subtle changes in the arrangement of water molecules in the surface layer. The frequency-independent relaxation contribution ranges from 2.43 s$^{-1}$ (313 K) to 4.60 s$^{-1}$ (263 K) for the dispersion of 25 nm nanoparticles and from 0.81 s$^{-1}$ (313 K) to 1.84 s$^{-1}$ (263 K) for the dispersion of 45 nm nanoparticles; in both cases, the changes in this temperature range are by a factor of about 2. This relaxation contribution is associated with dynamical processes of the order of tenths of nanoseconds or faster; the short correlation time implies that in the covered frequency range, one does not observe a frequency dependence. This term can describe the relaxation of the bulk-like fraction of water molecules in the dispersions, i.e., the fraction of water molecules, the dynamics of which is affected to a much lesser extent by the presence of the nanoparticles. One could expect that those water molecules perform isotropic translation motion, which is slowed down (compared to bulk water) by the presence of the nanoparticles, but the effect is much less significant than for the water molecules close to the surface and the movement remains three-dimensional. The slowing down is less significant for the dispersion of the 45 nm nanoparticles because of their lower concentration.

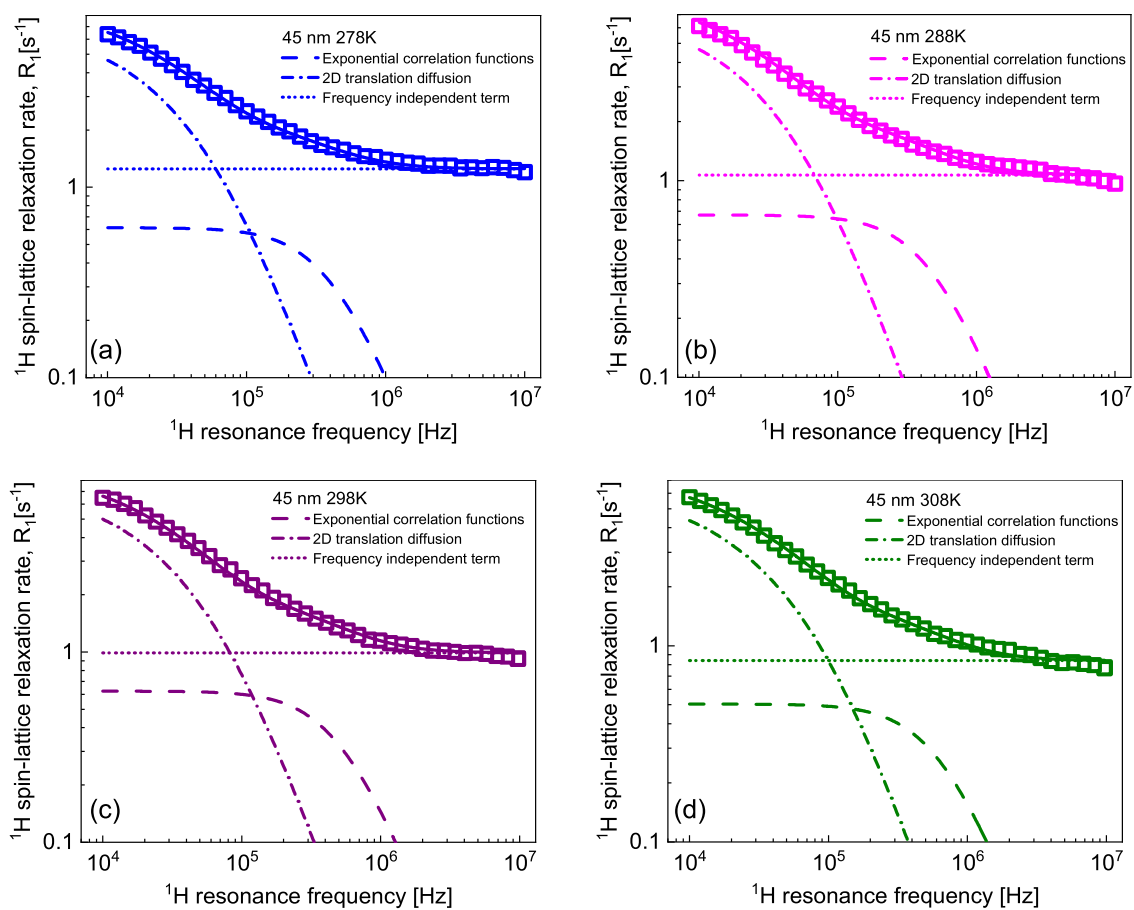

**Figure 4.** (a–d) $^1$H spin-lattice relaxation rates for water dispersion of silica nanoparticles of 45 nm diameter and $4.62 \times 10^{-3}$ mmol/dm$^3$ concentration. Solid lines, fits in terms of eq 6 decomposed into a relaxation contribution associated with two-dimensional translation diffusion (dashed-dotted lines), a relaxation contribution expressed in terms of Lorentzian spectral densities (dashed lines), and a frequency-independent term (dotted lines).

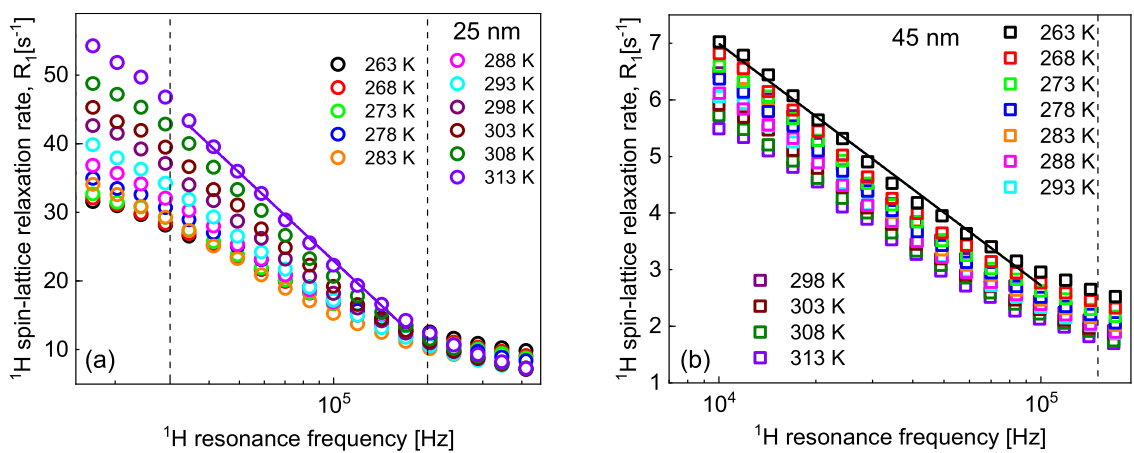

**Figure 5.** (a, b) $^1$H spin-lattice relaxation rates for water dispersions of silica nanoparticles functionalized with triethoxylpropylaminosilane; (a) 25 nm diameter and $2.44 \times 10^{-2}$ mmol/dm$^3$ and (b) 45 nm diameter and $4.62 \times 10^{-3}$ mmol/dm$^3$. Vertical lines indicate the frequency ranges in which linear dependencies of the relaxation rates on the logarithm of the resonance frequency are expected. Examples of the linear dependencies are shown for 313 and 263 K for 25 and 45 nm nanoparticles, respectively.

## VI. CONCLUSIONS

The thorough analysis of the $^1$H spin-lattice relaxation data has led to the conclusion that water molecules in the nanoparticle dispersions perform two-dimensional translation diffusion in a layer close to the nanoparticle surface. The diffusion process is mediated by acts of adsorption on the surface. The residence lifetime is temperature-independent and yields 4.54 × 10$^{-6}$ s for the smaller nanoparticles and 1.43 × 10$^{-5}$ s for the larger ones. The translation diffusion is also very weakly affected by temperature. For the smaller nanoparticles, the translational correlation time is, in fact, temperature-independent, 5.84 × 10$^{-7}$ s, while for the larger ones, it varies from 7.80 × 10$^{-7}$ to 1.18 × 10$^{-6}$ s (the ratio between the two values is about 1.5). The dynamics of water molecules attached to the surface of the smaller nanoparticles is characterized by a correlation time ranging from 4.86 × 10$^{-8}$ to 7.83 × 10$^{-8}$ s; the lower value is reached at 303 K and does not change with increasing temperature; analogously, the upper value is reached at 273 K and does not change with decreasing temperature. For the larger nanoparticles, the correlation time varies from 1.24 × 10$^{-7}$ to 2.19 × 10$^{-7}$ s (changing by about a factor of 2), reaching the last value already at 278 K. The scenario of the motion is complemented by the information that can be obtained from the dipolar relaxation constants, $C_{DD}$ and $C_{trans}$, suggesting that the fractions of water molecules that are bound to the nanoparticle surface and diffuse in the vicinity of the surface are lower for the larger nanoparticles, although one should be aware of exchange and molecular arrangement effects.

## ■ ASSOCIATED CONTENT

### Data Availability Statement

The data that support the findings of this study are available in 10.5281/zenodo.10256933.

■ Supporting Information

The Supporting Information is available free of charge at https://pubs.acs.org/doi/10.1021/acs.jpcb.3c06451.

Examples of $^1$H magnetization curves for water dispersions of silica nanoparticles (PDF)

■ AUTHOR INFORMATION

**Corresponding Author**

**Danuta Kruk** − *Department of Physics and Biophysics, University of Warmia & Mazury in Olsztyn, 10-719 Olsztyn, Poland;* orcid.org/0000-0003-3083-9395; Email: danuta.kruk@uwm.edu.pl

**Authors**

**Aleksandra Stankiewicz** − *Department of Physics and Biophysics, University of Warmia & Mazury in Olsztyn, 10-719 Olsztyn, Poland*

**Adam Kasparek** − *Department of Physics and Biophysics, University of Warmia & Mazury in Olsztyn, 10-719 Olsztyn, Poland;* orcid.org/0000-0002-3621-0881

**Elzbieta Masiewicz** − *Department of Physics and Biophysics, University of Warmia & Mazury in Olsztyn, 10-719 Olsztyn, Poland*

Complete contact information is available at:
https://pubs.acs.org/10.1021/acs.jpcb.3c06451

**Notes**
The authors declare no competing financial interest.

■ ACKNOWLEDGMENTS

This project has received funding from the European Union's Horizon 2020 Research and Innovation Program under Grant Agreement No. 899683 ("HIRES-MULTIDYN").